\texsis

\def\so{\overline s}

\def\ul#1{\underbar{#1}}

\paper
\titlepage
\hbox{\space}\vskip 1in
\title Particle Motion in the Stable Region Near the 
Edge of a Linear Sum Resonance Stopband
\endtitle
\author
George Parzen
October 23, 1995
BNL-62546
\endauthor
\abstract
This paper studies the particle motion when the tune is in the stable region
close to the edge of linear sum resonance stopband.  Results are found for
the tune and the beta functions.  Results are also found for the two
solutions of the equations of motion.  The results found are shown to be
also valid for small accelerators where the large accelerator approximation
may not be used.
\endabstract
\endtitlepage

\section{Introduction}

This paper studies the motion of a particle whose tune is near an edge of
a linear sum resonance stopband.  It is assumed that the tune is not near
any other linear resonance, and the motion is dominated by the linear
sum resonance.  It is assumed that the linear sum resonance is being
driven by a skew quadrupole field perturbation.  When the unperturbed
tune $\nu_{x0}, \nu_{y0}$ is close to the resonance line $\nu_x+\nu_y=q$,
$q$ being an integer, the particle motion can be unstable.  The region of
instability is called the stopband.  Results are found for the tune and
the beta functions when the unperturbed tune is in the stable region but
close to an edge of the stopband.  Results are also found for the two
solutions of the equations of motion.  All the results found are shown
to be also valid for small accelerators where the large accelerator
approximation may not be used.

\section{Results When The Tune Is Inside The Stopband}

It will be assumed that in the absence of the perturbing fields, the
tune of the particle is given by $\nu_{x0}, \nu_{y0}$, the $x$ and $y$
motions are uncoupled, and that the motion is stable when $\nu_{x0}, 
\nu_{y0}$ is close to the line $\nu_{x0}+\nu_{y0}=q$, where $q$ is an
integer.  It is assumed that a perturbing field is then added which is
given by the skew quadrupole field
$$\eqalign{
\Delta B_x &= - B_0 a_1 \ x \cr
\Delta B_y &= B_0 a_1 \ y \cr} \eqno{\hbox{(2-1)}}$$
$a_1$ is the skew quadrupole multipole and $a_1=a_1(s)$.  $B_0$ is some
standard field, usually the field in the main dipoles of the lattice.

The coupled equations of motion can be written as
$$\eqalign{
{d^2\eta_x\over d\theta_x^2} &+ \nu_{x0}^2 \eta_x = f_x \cr
{d^2\eta_y\over d\theta_y^2} &+ \nu_{y0}^2 \eta_y = f_y \cr
f_x &= \nu_{x0}^2 \beta_x^{3/2} \Delta B_y/B\rho \cr
f_y &= -\nu_{y0}^2 \beta_y^{3/2} \Delta B_x/B\rho \cr
\eta_x &= x/\beta_x^{1\over 2}, \quad \eta_y = y/\beta_y^{1\over 2} \cr
\rho &= B\rho/B_0 , \quad B\rho = pc/e \cr
ds &= \nu_{x0} \beta_x d\theta_x = \nu_{y0} \beta_y d\theta_y 
\cr} \eqno{\hbox{(2-2)}}$$
$\beta_x,\beta_y$ are the unperturbed beta functions.

Eqs. (2-2) are valid for large accelerators, and some changes are
required [1] to make them valid for small accelerators (see section 6).
However the final results found below are also valid for small accelerators
that require the use of the exact linearized equations.  This is shown
in section 6.

Eq. (2-2) can be rewritten as
$$\eqalign{
{d^2\over d\theta_x^2} \eta_x &+ \nu_{x0}^2 \eta_x = b_x \ \eta_y \cr
{d^2\over d\theta_y^2} \eta_y &+ \nu_{y0}^2 \eta_y = b_y \ \eta_x \cr
b_x &= -\nu_{x0}^2 \beta_x (\beta_x\beta_y)^{1\over 2} a_1/\rho \cr
b_y &= -\nu_{y0}^2 \beta_y (\beta_x\beta_y)^{1\over 2} a_1/\rho \cr}
\eqno{\hbox{(2-3)}}$$

Eqs. (2-3) are a set of linear equations for $\eta_x,\eta_y$ with
coefficients that are periodic in $s$.  The extension of Floquet's
theorem to more than one dimension applies, and the solutions have
the Floquet form,
$$\left[ \matrix{ \eta_x \cr \eta_y \cr} \right] = \exp(i\nu_x \theta_x)
\left[ \matrix{ h_x \cr h_y \cr} \right] \eqno{\hbox{(2-4)}}$$
$h_x$ and $h_y$ are periodic in $s$.  The solution can also have the form
$$\left[ \matrix{ \eta_x \cr \eta_y \cr} \right] = \exp(i\nu_y\theta_y)
\left[ \matrix{ h_x \cr h_y \cr} \right] \eqno{\hbox{(2-5)}}$$
If for small $a_1$ one finds a solution of the form Eq. (2-4) where
$h_y\rightarrow 0$ when $a_1\rightarrow 0$, then $\nu_x\rightarrow \nu_{x0}$.
This solution reduces to the uncoupled $x$ motion when $a_1\rightarrow 0$
and will be called the $\nu_x$ mode.  If for small $a_1$, one finds a
solution of the form Eq. (2-5) where $h_x\rightarrow 0$, when $a_1
\rightarrow 0$, then $\nu_y\rightarrow \nu_{y0}$ and thus this solution will
be called the $\nu_y$ mode.

To find the solution that corresponds to the $\nu_x$ mode, one can assume
that $\eta_x$ has the form
$$\eqalign{
\eta_x &= A_s \exp(i\nu_{xs}\theta_x) + \sum_{r\ne s} A_r \exp(i\nu_{xr}
\theta_x) \cr
\nu_{xr} &= \nu_{xs}+n,\quad n \hbox{\ an integer}, n\ne 0 \cr} 
\eqno{\hbox{(2-6a)}}$$
where for small enough $a_1$, $A_r\ll A_s$ and $\nu_{xs}\rightarrow\nu_{x0}$
for $a_1\rightarrow 0$.  For the corresponding form for $\eta_y$ one might
assume for $\eta_y$
$$\eqalign{
\eta_y &= \sum_r B_r \exp(i\nu_{yr}\theta_y) \cr
\nu_{yr} &= \nu_{xs} + n \cr} \eqno{\hbox{(2-6b)}}$$
where $B_r\ll A_s$ for small enough $a_1$.

Eqs. (2-6) have the form given by Eq. (2-4) for the $\nu_x$ mode.  It
will be seen below, that the solution assumed for $\eta_y$ Eq. (2-6b)
is valid if one is not near the sum resonance $\nu_x+\nu_y=q$, $q$ being
an integer.  When $\nu_{x0},\nu_{y0}$ are close to the sum resonance
$\nu_x+\nu_y=q$, then one of the $B_r$ will become as large as $A_s$ and
this is the $B_r$ for which $\nu_{yr}=\nu_{xs}-q$.  This is shown below.
Thus, one assumes for $\eta_y$ the solution with the form
$$\eqalign{
\eta_y &= B_{\so}\exp(i\nu_{y\so}\theta_y)+\sum_{r\ne\so} B_r \exp(i\nu_{yr}
\theta_y) \cr
\nu_{y\so} &= \nu_{xs} - q \cr
\nu_{yr} &= \nu_{xs} + n, \quad n\ne -q \cr} \eqno{\hbox{(2-6c)}}$$
Here $B_r\ll A_s$ but $B_{\so}\simeq A_s$.  It is being assumed that
$\nu_{x0}$, $\nu_{y0}$ are not close to any other resonance other than
$\nu_x+\nu_y=q$.

Putting this assumed form for $\eta_x$, $\eta_y$ into the differential
equations Eq. (2-3), one gets the equations for $A_r$, $B_r$
$$\eqalign{
(\nu_{xr}^2-\nu_{x0}^2)A_r &= -2\nu_{x0}\sum_{r'} b_x(-\nu_{xr},\nu_{yr'})
B_{r'} \cr
(\nu_{yr}^2-\nu_{y0}^2)B_r &= -2\nu_{y0}\sum_{r'} b_y(-\nu_{yr},\nu_{xr'})
A_{r'} \cr
b_x(-\nu_{xr},\nu_{yr'}) &= {1\over 4\pi} \int_0^L ds(\beta_x\beta_y)^{1\over 2}(a_1/\rho)\exp[i(-\nu_{xr}\theta_x+\nu_{yr'}\theta_y)] \cr
b_y(-\nu_{yr},\nu_{xr'}) &= {1\over 4\pi} \int_0^L ds(\beta_x\beta_y)^{1\over 2}(a_1/\rho)\exp[i(-\nu_{yr}\theta_y+\nu_{xr'}\theta_x)] \cr} \eqno{\hbox{(2-7)}}$$
$L$ is the lattice circumference.

Eqs. (2-7) can be solved by an iterative perturbation procedure.  For the
initial guess for $\eta_x$, $\eta_y$ in the iterative procedure one can
assume
$$\eqalign{
\eta_x &= A_s \exp(i\nu_x\theta_x) \cr
\eta_y &= B_{\so}\exp(i\nu_{y\so}\theta_y) \cr
\nu_{y\so} &= \nu_{xs}-q=-(q-\nu_{xs}) \cr} \eqno{\hbox{(2-8)}}$$
One can put this initial guess for $\eta_x$, $\eta_y$ in the right hand
side of Eq. (2-7) and solve for the $A_r$, $B_r$ which gives
$$\eqalign{
(\nu_{xr}^2-\nu_{x0}^2) A_r &= -2\nu_{x0} b_x(-\nu_{xr},\nu_{y\so}) B_{\so} \cr
(\nu_{yr}^2-\nu_{y0}^2) B_r &= -2\nu_{y0} b_y(-\nu_{yr},\nu_{xs}) A_s \cr
\nu_{xr} &= \nu_{xs} + n , \quad \nu_{yr}=\nu_{xs}+m \cr} \eqno{\hbox{(2-9)}}$$
For $A_r=A_s$ and $B_r=B_{\so}$ one finds
$$\eqalign{
(\nu_{xs}^2-\nu_{x0}^2)A_s &= -2\nu_{x0}b_x(-\nu_{xs},\nu_{y\so}) B_{\so} \cr
(\nu_{y\so}^2-\nu_{y0}^2)B_{\so} &= -2\nu_{y0}b_x(-\nu_{y\so},\nu_{xs}) A_s \cr
\nu_{y\so} &= - (q-\nu_{xs}) \cr} \eqno{\hbox{(2-10)}}$$
Eqs. (2-10) are two linear homogeneous equations for $A_s$ and $B_{\so}$, and
to be solvable, we must have
$$\eqalign{
(\nu_{xs}^2-\nu_{x0}^2) (\nu_{y\so}^2-\nu_{y0}^2) &= 4\nu_{x0}\nu_{y0}
|\Delta\nu_x|^2 \cr
\Delta\nu_x &= {1\over 4\pi} \int_0^L ds (\beta_x\beta_y)^{1\over 2} (a_1/\rho)
\exp[-i(\nu_{x0}\theta_x+(q-\nu_{x0})\theta_y)] \cr} \eqno{\hbox{(2-11)}}$$
where one uses $b_y(-\nu_{y\so},\nu_{xs})=b_x^*(-\nu_{xs},\nu_{y\so})$.  Eq.
(2-11) is an equation for $\nu_{xs}$, the tune of the $\nu_x$ mode, which
is the mode where $\nu_{xs}\rightarrow\nu_{x0}$ when $a_1\rightarrow 0$.
It will be assumed that $\nu_{x0}$, $\nu_{y0}$ is close to the resonance
line $\nu_x+\nu_y=q$ and one can write
$$\eqalign{
(\nu_{xs}^2-\nu_{x0}^2) &= (\nu_{xs} + \nu_{x0}) (\nu_{xs}-\nu_{x0}) \simeq
2\nu_{x0} (\nu_{xs}-\nu_{x0}) \cr
(\nu_{y\so}^2-\nu_{y0}^2) &= (\nu_{y\so}+\nu_{y0}) (|\nu_{y\so}|-
\nu_{y0}) \simeq 2\nu_{y0} (q-\nu_{xs}-\nu_{y0}) \cr} \eqno{\hbox{(2-12)}}$$
Eq. (2-11) then becomes
$$(\nu_{xs}-\nu_{x0}) (q-\nu_{xs}-\nu_{y0}) = |\Delta\nu_x| 
\eqno{\hbox{(2-13)}}$$
To solve Eq. (2-13) one puts
$$\nu_{xs} = \nu_{xsR}-ig_x$$
where $\nu_{xsR}$ and $g_x$ are both real, which gives the equation
$$(\nu_{xsR}-ig_x-\nu_{x0}) (q-\nu_{xsR}+ig_x-\nu_{y0})=|\Delta\nu_x|^2
\eqno{\hbox{(2-14)}}$$
The imagninary part of Eq. (2-14) gives
$$g_x[\nu_{xsR}-\nu_{x0} (q-\nu_{xsR}-\nu_{y0})]=0 \eqno{\hbox{(2-15)}}$$
If one is inside the stopband, then $g_x\ne 0$ and one gets
$$\nu_{xsR} = {1\over 2} [\nu_{x0}+q-\nu_{y0}] \eqno{\hbox{(2-16)}}$$
The real part of Eq. (2-14) gives
$$(\nu_{xsR}-\nu_{x0})(q-\nu_{xsR}-\nu_{y0})+q_x^2=|\Delta\nu_x|^2
\eqno{\hbox{(2-17)}}$$
Using Eq. (2-16) for $\nu_{xsR}$ one has
$$\eqalign{
\nu_{xsR}-\nu_{x0} &= {1\over 2} [-\nu_{x0}+(q-\nu_{y0})] \cr
q-\nu_{xsR}-\nu_{y0} &= {1\over 2} [(q-\nu_{y0})-\nu_{x0}] \cr}
\eqno{\hbox{(2-18)}}$$
Eq. (2-17) then gives
$$\eqalign{
g_x^2 &+ [{1\over 2} (q-\nu_{x0}-\nu_{y0})]^2 = |\Delta\nu_x|^2 \cr
g_x &= \pm \left\{ |\Delta\nu_x|^2 - [{1\over 2}(q-\nu_{x0}-\nu_{y0})]^2
\right\}^{1/2} \cr} \eqno{\hbox{(2-19)}}$$
Eq. (2-19) shows that the growth factor $g_x$ has a maximum value of
$g_x=|\Delta\nu_x|$ when $\nu_{x0}$, $\nu_{y0}$ are on the resonance line,
$q-\nu_{x0}-\nu_{y0}=0$, and then decreases to zero at the edges of the
stopband given by the two lines
$$q-\nu_{x0}-\nu_{y0}=\pm 2|\Delta\nu_x| \eqno{\hbox{(2-20)}}$$
Eq. (2-19) shows that the unstable region in $\nu_{x0}$, $\nu_{y0}$ is
bounded by the two lines given by Eq. (2-20).  These two lines are
parallel to the resonance line $q-\nu_{x0}-\nu_{y0}=0$, which lies
midway between these two lines.  If one wanted to define a stopband
width, one might define it as the distance in $\nu_{x0}$, $\nu_{y0}$
space across the unstable region, along a path which is perpendicular
to the two boundary lines.  This is given by
$$\hbox{stopband\ width} = 2.828 |\Delta\nu_x| \eqno{\hbox{(2-21)}}$$

For particle motion in 2 dimensional phase space, it has been found [2]
that the real part of the tune is constant as the unperturbed tune moves
across the stopband.  This is not in general true for 4 dimensional phase,
as the real part of the tune, given by Eq. (2-16), depends on the path
in $\nu_{x0}$, $\nu_{y0}$ which is chosen in crossing the stopband.
However, if one chooses a path which is perpendicular to the resonance
line, $q-\nu_{x0}-\nu_{y0}=0$, then the real part of the tune does 
remain constant.  One can see this by observing that if starting from
$\nu_{x0}$, $\nu_{y0}$ one draws a line perpendicular to the resonance
line, the point on the resonance line that this perpendicular meets has
the coordinates
$${1\over 2} (\nu_{x0}+q-\nu_{y0}), \ \ {1\over 2} (\nu_{y0}+q-\nu_{x0})
\eqno{\hbox{(2-22)}}$$
The $\nu_x$ coordinate of this point is just $\nu_{xsR}$ as given by
Eq. (2-16).

The above results are for the $\nu_x$ mode, the mode for which the tune
approaches $\nu_{x0}$ when $a_1\rightarrow 0$.  One can find the
corresponding results for the $\nu_y$ mode by using the following
substitutions
$$\eqalign{
\nu_{x0} &\rightarrow \nu_{y0} \cr
\nu_{y0} &\rightarrow \nu_{x0} \cr
\Delta\nu_x &\rightarrow \Delta\nu_y \cr
g_x &\rightarrow g_y \cr} \eqno{\hbox{(2-23)}}$$
The apparent differences between $g_x$ and $g_y$ and $\Delta\nu_x$ and
$\Delta\nu_y$ are negligible if the $\nu_{x0}$, $\nu_{y0}$ are close to
the resonance line.  $\Delta\nu_x$, $\Delta\nu_y$ can be written as
$$\Delta\nu={1\over 4\pi} \int ds (\beta_x\beta_y)^{1\over 2} (a_1/\rho)
\exp[-i(\nu_x\theta_x+\nu_y\theta_y)] \eqno{\hbox{(2-24a)}}$$
where for the $\nu_x$ mode, $\nu_x$, $\nu_y$ is the point on the
resonance line
$$\nu_x=\nu_{x0}, \nu_y=q-\nu_{x0} \eqno{\hbox{(2-24b)}}$$
and for the $\nu_y$ mode, $\nu_x$, $\nu_y$ is the point on the
resonance line.
$$\nu_x=q-\nu_{y0}, \nu_y=\nu_{y0} \eqno{\hbox{(2-24c)}}$$
These two points on the resonance line are close if $\nu_{x0}$, $\nu_{y0}$
is assumed to be close to the resonance line, and their difference can
be neglected.  A reasonable compromise might be to choose for $\nu_x$,
$\nu_y$ the point on the resonance line which is midway between these two
points, which is the choice of $\nu_x$, $\nu_y$ given by
$$\nu_x={1\over 2} (\nu_{x0}+q-\nu_{y0}), \nu_y = {1\over 2} (\nu_{y0} +
q-\nu_{x0}) \eqno{\hbox{(2-25)}}$$

\section{Solutions of the Equations of Motion}

Now let us find the solutions for $\eta_x$, $\eta_y$ that will give the
particle motion inside the stopband.  To lowest order $\eta_x$ and
$\eta_y$ are given by Eqs. (2-6) for the $\nu_x$ mode as
$$\eqalign{
\eta_x &= A_s \exp (i\nu_{xs}\theta_x) \cr
\eta_y &= B_{\so} \exp(i\nu_{y\so}\theta_y) \cr
\nu_{y\so} &= \nu_{xs}-q \cr} \eqno{\hbox{(3-1)}}$$
From Eq. (2-10) one finds that inside the stopband
$$B_{\so} = - {\Delta\nu_x^*\over |\nu_{y\so}|-\nu_{y0}} A_s
\eqno{\hbox{(3-2)}}$$
Since
$$\eqalign{
|\nu_{y\so}| - \nu_{y0} &= q-\nu_{xs}-\nu_{y0} \cr
&= q-\nu_{xsR}+ig_x-\nu_{y0} \cr
&= q-\nu_{y0}-{1\over 2} [\nu_{x0}+q-\nu_{y0}]+ig_x \cr
&= {1\over 2} (q-\nu_{x0}-\nu_{y0})+ig_x \cr} \eqno{\hbox{(3-3)}}$$
one gets
$$\eqalign{
B_{\so} &= -\exp [-i(\delta_{1x}+\delta_{2x})] A_s \cr
\delta_{1x} &= ph (\Delta\nu_x) \cr
\delta_{2x} &= ph [{1\over 2} (q-\nu_{x0}-\nu_{y0})+ig_x] \cr} 
\eqno{\hbox{(3-4)}}$$
where use was made of Eq. (2-19) and $ph$ indicates the phase of a
variable.

Thus, inside the stopband, $\eta_x$ and $\eta_y$ are of the same order of
magnitude one can find the first order correction to $\eta_x$ and $\eta_y$
using Eq. (2-9).  The results for $\eta_x$ and $\eta_y$ for the $\nu_y$
mode can be found by using the substitutions given by Eqs. (2-23).

Results will now be found for $\eta_x$ and $\eta_y$ which are correct to
first order in the perturbation and when $\nu_{x0}$, $\nu_{y0}$ is inside
the stopband or in the stable region near an edge of the stopband.
$\eta_x$, $\eta_y$ are given by Eqs. (2-6).  For the $\nu_x$ mode
$$\eqalign{
\eta_x &= A_s \exp (i\nu_{xs}\theta_x)+\sum_{r\ne s} A_r \exp (i\nu_{xr}
\theta_x) \cr
\eta_y &= B_{\so} \exp(i\nu_{y\so}\theta_y)+\sum_{r\ne s} B_r \exp
(i\nu_{yr}\theta_y) \cr
\nu_{y\so} &= \nu_{xs} - q \cr
\nu_{yr} &= \nu_{xs} + n, \ \ n\ne-q \cr
\nu_{xr} &= \nu_{xs}+n,\ \ n\ne 0 \cr} \eqno{\hbox{(3-5)}}$$
The first order solution is given by Eqs. (2-9) and (2-10).  $B_{\so}$,
$A_r$ and $B_r$ are given by Eq. (2-7)
$$\eqalign{
B_{\so} &= -2\nu_{y0} b_y(-\nu_{y\so}, \nu_{xs}) A_s/(\nu_{y\so}^2-
\nu_{y0}^2) \cr
A_r &= -2\nu_{x0}b_x(-\nu_{xr},\nu_{y\so}) B_{\so}/(\nu_{xr}^2-\nu_{x0}^2) \cr
B_r &= -2\nu_{y0} b_y(-\nu_{yr},\nu_{xs}) A_s/(\nu_{yr}^2-\nu_{y0}^2) \cr
\nu_{xr} &= \nu_{xs} - m \ \ \ m\ne 0 \cr
\nu_{yr} &= \nu_{y\so}+n\ \ \ n\ne 0 \cr} \eqno{\hbox{(3-6)}}$$
First, let us compute $B_{\so}$.
$$\eqalign{
\nu_{y\so}^2 - \nu_{y0}^2 &= (|\nu_{y\so}|+\nu_{y0})(|\nu_{y\so}|-\nu_{y0}) \cr
&= 2\nu_{y0} (q-\nu_{xs}-\nu_{y0}) \cr} \eqno{\hbox{(3-7)}}$$
We can write for $\nu_{xs}$
$$\nu_{xs} = {1\over 2} (\nu_{x0}+q-\nu_{y0})-\delta_x \eqno{\hbox{(3-8)}}$$
where $\delta_x=ig_x$, see Eq. (2-19) for $g_x$, when the tune is inside the
stopband.  In the stable region near an edge of a stopband, $\nu_{xs}$ and
$\delta_x$ are given in section 4.  Note that $\delta_x=0$ when the tune
is on the edge of the stopband.  Then Eq. (3-7) gives
$$\nu_{y\so}^2-\nu_{y0}^2=2\nu_{y0}({1\over 2}(q-\nu_{x0}-\nu_{y0}) +
\delta_x) \eqno{\hbox{(3-9)}}$$
One also finds
$$\eqalign{
b_x (-\nu_{y\so},\nu_{xs}) &= \Delta\nu_x^* \cr
B_{\so} &= -d_x \exp(-i\delta_{1x}) A_s \cr
d_x &= - {|\Delta\nu_x|\over{1\over 2}(q-\nu_{x0}-\nu_{y0})+\delta_x}
\exp(-i\delta_{1x}) A_s \cr
\delta_{1x} &= ph (\Delta\nu_x) \cr} \eqno{\hbox{(3-10)}}$$
One may note that inside the stopband
$$\eqalign{
d_x &= -\exp[-i(\delta_{1x}+\delta_{2x})] A_s \cr
\delta_{2x} &= ph [{1\over 2}(q-\nu_{x0}-\nu_{y0})+ig_x] \cr}
\eqno{\hbox{(3-11)}}$$
One may now find the $A_r$ from Eq. (3-6)
$$\eqalign{
\nu_{xr} &= \nu_{xs}-m, \qquad m\ne 0 \cr
\nu_{xr}^2-\nu_{x0}^2 &= m (m-2\nu_{x0}) \cr
b_x(-\nu_{xr},\nu_{y\so}) &= b_m \cr
b_m &= {1\over 4\pi} \int ds ({a_1\over\rho})(\beta_x\beta_y)^{1\over 2}
\exp[-i((q-\nu_{x0})\theta_y+\nu_{x0}\theta_x)+im\theta_x] \cr
A_r &= {-2\nu_{x0}\over m(m-2\nu_{x0})} b_m B_{\so} \cr
A_r &= {-2\nu_{x0}\over m(m-2\nu_{x0})} d_x b_m \exp (-i\delta_{1x}) A_s
\cr} \eqno{\hbox{(3-12)}}$$
One may find the $B_r$ from Eq. (3-6)
$$\eqalign{
\nu_{yr} &= \nu_{y\so}+n, \qquad n=0 \cr
\nu_{y\so} &= \nu_{xs}-q \cr
\nu_{yr}^2-\nu_{y0}^2 &= n(n-2(q-\nu_{x0})) \cr
b_y(-\nu_{xr}, \nu_{xs}) &= c_n^* \cr
c_n &= {1\over 4\pi} \int ds ({a_1\over\rho})(\beta_x\beta_y)^{1\over 2}
\exp[-i((q-\nu_{x0})\theta_y+\nu_{x0}\theta_x)+in\theta_y] \cr
B_r &= {-2\nu_{y0}\over n(n-2(q-\nu_{x0}))} c_m^*\ A_s \cr} 
\eqno{\hbox{(3-13)}}$$
Putting these results for $B_{\so}$, $A_r$, $B_r$ into Eq. (3-5), one
finds the following results for $\eta_x$, $\eta_y$ for the $\nu_x$
mode.
$$\eqalign{
\eta_x &= A_s \exp(i\nu_{xs}\theta_x)[1+\sum_{m\ne 0} f_m\exp(-im\theta_x)] \cr
f_m &= {-2\nu_{x0}\over m(m-2\nu_{x0})} d_x b_m \exp (-i\delta_{1x}) \cr
d_x &= {-|\Delta\nu_x|\over{1\over 2}(q-\nu_{y0}-\nu_{x0})+\delta_x} \cr
\eta_y &= A_s \exp (i\nu_{y\so}\theta_y)[1+\sum_{n\ne 0} g_n \exp(in\theta_y)] \cr
g_n &= {-2\nu_{y0}\over n(n-2(q-\nu_{x0}))} C_n^* \cr
\nu_{y\so} &= \nu_{xs}-q \cr} \eqno{\hbox{(3-14)}}$$
$\delta_x$ is given by $i\ g_x$ inside the stopband where $g_x$ is given
by Eq. (2-19) as
$$g_x=\left\{|\Delta\nu_x|^2-[{1\over 2}(q-\nu_{x0}-\nu_{y0})]^2\right\}^{1\over 2} 
\eqno{\hbox{(3-15a)}}$$
In the stable region near a stopband edge, $\delta_x$ is given by Eq. (4-10)
as
$$\eqalign{ 
|\delta_x| &= \left\{ \epsilon_x (|\Delta\nu_x|+\epsilon_x/4)\right\}^{1\over 2} \cr
\epsilon_x &= |q\pm|2\Delta\nu_x|-\nu_{x0}-\nu_{y0}| \cr} \eqno{\hbox{(3-15b)}}$$
One uses the $+$ sign for the upper stopband edge and the $-$ sign for the
lower edge.  $\delta_x$ is positive for the lower edge and negative for the
upper edge.

Equations (3-4) give the solutions of the equations of motion for the
$\nu_x$ mode.  The solutions for the $\nu_y$ mode are found by replacing
each parameter for the $\nu_x$ mode by its corresponding parameter for the
$\nu_y$ mode.

One may note from Eq. (3-14) that for the $\nu_x$ mode the dominant harmonic
for $\eta_x$ is $m\simeq 2\nu_{x0}$, and for $\eta_y$ $n=2|\nu_{y\so}|=2(q-
\nu_{x0})$.  Keeping just the dominant harmonics give fairly simple results
for $\eta_x$, $\eta_y$.

\section{The Tune Near the Edge of a Stopband}

In this section, a result will be found for the tune in the stable region
outside the stopband but close to an edge of the stopband.  It will be
shown that close to an edge of the stopband the tune of the $\nu_x$ mode is
given by
$$\eqalign{
|\nu_x-{1\over 2} (\nu_{x0}+q-\nu_{y0})| &= \left\{\epsilon_x|\Delta\nu_x|\right\}^{1\over 2} \cr
\epsilon_x &= |q\pm 2|\Delta\nu_x|-\nu_{x0}-\nu_{y0}| \cr} \eqno{\hbox{(4-1)}}$$
$\nu_x$ is the tune of the $\nu_x$ mode, $\epsilon$ is the distance from
$\nu_{x0}$, $\nu_{y0}$ to the edge of the stopband.  In the $\pm$, the $+$
sign is for the upper edge, and the $-$ sign for the lower edge.  When
$\nu_{x0}$, $\nu_{y0}$ reaches the edge of the stopband, then $\epsilon=0$,
and $\nu_x={1\over 2}(\nu_{x0}+q-\nu_{y0})$, which according to Eq. (2-16)
is the real part of the tune inside the stopband.

Eq. (4-1) shows that near the stopband edge, $\nu_x$ varies rapidly with
$\epsilon_x$.  As one reaches the edge of the stopband, $\epsilon_x$ goes
to zero and $d\nu_x/d\epsilon_x$ becomes infinite like $\epsilon_x^{-{1\over 2}}$.

To find $\nu_x$ in the stable region outside the stopband, where
$|q-\nu_{x0}-\nu_{y0}|>2|\Delta\nu_x|$, one goes back to the derivation
given in section 2 for $\nu_x$ inside the stopband, starting with Eq. (2-13)
$$(\nu_x-\nu_{x0})(q-\nu_x-\nu_{y0})=|\Delta\nu_x|^2 \eqno{\hbox{(4-2)}}$$
Because of the condition that $\nu_x$ is outside the stopband or
$$|q-\nu_{x0}-\nu_{y0}|>2|\Delta\nu_x| \eqno{\hbox{(4-3)}}$$
one sees that one must have $g_x=0$, as Eq. (2-19) would indicate $g_x$
is imaginary which contradicts the assumption that $g_x$ is real.

Let us assume that we start with $\nu_{x0}$, $\nu_{y0}$ below the lower
stopband edge and let $\nu_{x0}$, $\nu_{y0}$ approach the lower stopband
edge.  The equation of the lower stopband edge is given by
$$q-\nu_{x0}-\nu_{y0}=2|\Delta\nu_x| \eqno{\hbox{(4-4)}}$$
when $\nu_{x0}$, $\nu_{y0}$ arrive on the lower stopband edge, then $\nu_x$
will arrive at the value $\nu_x={1\over 2}(\nu_{x0}+q-\nu_{y0})$ as 
indicated by Eq. (2-16).  Thus below the stopband edge one can write
$$\nu_x={1\over 2} (\nu_{x0}+q-\nu_{y0})-\delta_x \eqno{\hbox{(4-5)}}$$
where $\delta_x\rightarrow 0$ when $\nu_{x0}$, $\nu_{y0}$ arrive at the
stopband edge.  We then find
$$\eqalign{
\nu_x-\nu_{x0} &= {1\over 2} (q-\nu_{x0}-\nu_{y0})-\delta_x \cr
q-\nu_x-\nu_{y0} &= {1\over 2} (q-\nu_{x0}-\nu_{y0})+\delta_x \cr}
\eqno{\hbox{(4-6)}}$$
and Eq. (3-2) becomes
$$\eqalign{
&\phantom{} [{1\over 2}(q-\nu_{x0}-\nu_{y0})]^2 - \delta_x^2 = |\Delta\nu_x|^2 \cr
\delta_x &= \left\{[{1\over 2}(q-\nu_{x0}-\nu_{y0})]^2-|\Delta\nu_x|^2
\right\}^{1\over 2} \cr} \eqno{\hbox{(4-7)}}$$
Eq. (4-7) gives $\nu_x$ in the stable region near the stopband.  It can be
put in another form that indicates the dependence on the distance from
$\nu_{x0}$, $\nu_{y0}$ to the stopband edge.

Below the stopband, one writes
$$\epsilon_x=q-2|\Delta\nu_x|-\nu_{x0}-\nu_{y0} \eqno{\hbox{(4-8)}}$$
where $\epsilon_x$ indicates the distance from $\nu_{x0}$, $\nu_{y0}$ to
the stopband edge which is given by Eq. (4-4).  When $\nu_{x0}$, $\nu_{y0}$
is on the stopband edge and $\nu_{x0}+\nu_{y0}=q-2|\Delta\nu_x|$ then
$\epsilon_x=0$.

Using Eq. (4-8) to replace $q-\nu_{x0}-\nu_{y0}$ by $\epsilon_x+2|\Delta\nu_x|$
in Eq. (3-7) one finds
$$\delta_x=\left\{\epsilon_x(|\Delta\nu_x|+\epsilon_x/4)\right\}^{1\over 2} 
\eqno{\hbox{(4-9)}}$$
Eq. (4-9) can then be written so as to hold both above and below the
stopband to give
$$\eqalign{
\left|\nu_x-{1\over 2}(\nu_{x0}+q-\nu_{y0})\right| &= \left\{\epsilon_x
(|\Delta\nu_x|+\epsilon_x/4)\right\}^{1\over 2} \cr
\epsilon_x &= |q\pm2|\Delta\nu_x|-\nu_{x0}-\nu_{y0}| \cr} 
\eqno{\hbox{(4-10)}}$$
where $\epsilon_x$ is the distance from $\nu_{x0}$, $\nu_{y0}$ to the
stopband edge.  One uses the $+$ sign for the upper stopband edge and the
$-$ sign for the lower edge.

Close to the stopband edge, where $\epsilon_x\ll |\Delta\nu_x|$ then
Eq. (4-10) gives the result
$$\left| \nu_x-{1\over 2} (\nu_{x0}+q-\nu_{y0})\right|= \left\{ \epsilon_x
\left| \Delta\nu_x \right| \right\}^{1/2} \eqno{\hbox{(4-11)}}$$
Equations (4-10) and (4-11) give the tune of the $\nu_x$ mode, $\nu_x$,
near the stopband edge.  The result for the tune of the $\nu_y$ mode,
$\nu_y$, may be found by making the substitution $\nu_x\rightarrow \nu_y$,
$\nu_{x0}\rightarrow\nu_{y0}$, $\nu_{y0}\rightarrow\nu_{x0}$, $|\Delta\nu_x|
\rightarrow |\Delta\nu_y|$.

If one varies the unperturbed tune, $\nu_{x0}$, $\nu_{y0}$, so that the
tune approaches the edge of the stopband, the tune on the stopband edge
depends on the value of $\nu_{x0}$, $\nu_{y0}$ when the unperturbed tune
arrives at the stopband edge.  The stopband edges are given by the two lines
$$\nu_{x0}+\nu_{y0}=q\pm 2|\Delta\nu|$$
where it is assumed that $|\Delta\nu_x|=|\Delta\nu_y|=|\Delta\nu|$ and the
$+$ sign is for the upper edge and the $-$ sign for the lower edge.

The tune of the $\nu_x$ mode at the stopband edge is then given by
$$\eqalign{
\nu_x &= {1\over 2} (\nu_{x0}+q-\nu_{y0}) \cr
\nu_x &= \nu_{x0}\pm |\Delta\nu| \cr} \eqno{\hbox{(4-12)}}$$
where the $+$ sign is for the lower edge and the $-$ sign for the upper
edge.

The tune of the $\nu_y$ mode at the stopband edge is given by
$$\nu_y = \nu_{y0} \pm |\Delta\nu|$$
One may note, that at the stopband edge
$$\eqalign{
\nu_x+\nu_y &= \nu_{x0}+\nu_{y0}\pm 2|\Delta\nu| \cr
\nu_x+\nu_y &= q \cr} \eqno{\hbox{(4-13)}}$$
and the $\nu_x$, $\nu_y$ lies on the resonance line.

Eqs. (4-6) and (4-7) can also be rewritten as, for the $\nu_x$ mode and
below the resonance line,
$$\eqalign{
\nu_x &= \nu_{x0} + 0.5 D \left\{ 1-[({2\Delta\nu\over D})^2]^{1\over 2}
\right\} \cr
D &= q-\nu_{x0}-\nu_{y0} \cr
\Delta\nu &= \Delta\nu_x \simeq \Delta\nu_y \cr} \eqno{\hbox{(4-14a)}}$$
The equation for the $\nu_y$ mode is similar
$$\nu_y = \nu_{y0}+0.5 D\left\{1-[({2\Delta\nu\over D})^2]^{1\over 2}
\right\} \eqno{\hbox{(4-14b)}}$$
Using Eq. (4-14) one can compute how much $\nu_x$, $\nu_y$ will move
given the distance from the resonance, $D$, and the stopband width
$\Delta\nu$.  Eq. (4-14) show that as $\Delta\nu$ is increased, $\nu_x$
and $\nu_y$ will move along the line from $\nu_{x0}$, $\nu_{y0}$ which
is perpendicular to the resonance line.  When $\Delta\nu$ reaches
0.5 $D$, $\nu_x$, $\nu_y$ will arrive on the resonance line where
$\nu_x=\nu_{x0}+0.5D$, $\nu_y=\nu_{y0}+0.5D$, $\nu_x+\nu_y=q$.

\section{The Beta Functions Near the Edge of a Stopband}

In this paper it is being assumed that the unperturbed tune $\nu_{x0}$,
$\nu_{y0}$ is near the sum resonance $\nu_{x0}+\nu_{y0}=q$, $q$ an 
integer, and the other linear resonances are far enough away so that the
particle motion is dominated by the sum resonance.  For this case, it will
be shown that the beta functions $\beta_x$, $\beta_y$ do not become
infinite when $\nu_{x0}$, $\nu_{y0}$ approach the edge of the stopband,
as was found [2] for the case of uncoupled particle motion near a half
integer resonance.  If one goes to a coordinate system where the
coordinates are uncoupled, then $\beta_x$ for the uncoupled coordinates
can become infinite, when the tune of the $\nu_x$ mode, $\nu_x$ is close
to a half integer resonance, $\nu_x\simeq n/2$, $n$ being an integer.
It is assumed here that $\nu_{x0}$, and thus $\nu_x$ is not near a 
half--integer resonance.

The beta functions, $\beta_x$, $\beta_y$, for linearly coupled motion may
be defined by going to the coordinate system where the new coordinates
$u$, $p_u$, $v$, $p_v$ are uncoupled.  If the $u$, $p_u$ motion goes
over into $x$, $p_x$ motion, when the coupling goes to zero, then the
beta function of the $u$, $p_u$ motion will be called $\beta_x$.  A
similar definition is given to $\beta_y$.

In section 2, a solution was found for $\eta_x$, and $x=\beta_{x0}^{1\over 2}
\eta_x$, when $\nu_{x0}$, $\nu_{y0}$ are in the stable region near the
edge of the stopband.  $\beta_x$ can be computed from this solution using
the result, see section 6,
$${1\over\beta_x} = {1\over\nu_{x0}\beta_{x0}} Im {d\over d\theta_x} 
\log x \eqno{\hbox{(5-1)}}$$
$Im$ stand for the imaginary part and $\nu_{x0}$, $\beta_{x0}$ are the 
tune and beta function of the unperturbed motion.  Eq. (5-1) holds for
large accelerators.  For small accelerators, where the large accelerator
approximations are not used, it also requires that $B_x=0$ on the closed
orbit.  Eq. (5-1) is derived in section 6.

The change in $\beta_x$ due to the linear coupling field may be computed
using Eq. (5-1) and the solution for $x=\beta_{x0}^{1\over 2}\eta_x$ found
in section 3, Eq. (3-14),
$$\eqalign{
\eta_x &= A_s \exp(i\nu_{x0}\theta_x) [1+\sum_{m\ne 0} f_m \exp(-im\theta_x)]\cr
f_m &= {-2\nu_{x0}\over m(m-2\nu_{x0})} d_x b_m \exp (-i\delta_{1x}) \cr
d_x &= {-|\Delta\nu_x|\over{1\over 2}(q-\nu_{x0}-\nu_{y0})+\delta_x} \cr
b_m &= {1\over 4\pi} \int ds {a_1\over\rho} (\beta_x\beta_y)^{1/2} \exp
[-i((q-\nu_{x0})\theta_y+\nu_{x0}\theta_x)+im\theta_x] \cr
\Delta\nu_x &= b_0, \quad \delta_{1x} = ph (\Delta\nu_x) \cr}
\eqno{\hbox{(5-2)}}$$
In the stable region, near a stopband edge, $\delta_x$ is given by
$$\eqalign{
|\delta_x| &= \left\{ \epsilon(|\Delta\nu_x|+\epsilon/4) \right\}^{1\over 2}\cr
\epsilon &= |q+\pm|2\Delta\nu_x|-\nu_{x0}-\nu_{y0}| \cr} \eqno{\hbox{(5-3)}}$$
with the $\pm$ sign for the upper and lower edge, respectively.  $\nu_{xs}$
has been replaced by $\nu_x$ and $\delta_x$ is positive for the lower edge
and negative for the upper edge.  One then gets
$$\eqalign{
Im {d\over d\theta_x} \log x &= \nu_x+Im\sum_{m\ne 0} (-imf_m\exp(-im\theta_x)) \cr
{1\over\beta_x} &= {1\over\nu_{x0}\beta_{x0}} (\nu_x-Re\sum_{m\ne 0} m f_m
\exp(-im\theta_x)) \cr
{1\over\beta_x} &= {1\over\beta_{x0}} (1+{\nu_x-\nu_{x0}\over\nu_{x0}}-
{1\over\nu_{x0}} Re \sum_{m\ne 0} m f_m \exp(-im\theta_x)) \cr
\beta_x &= \beta_{x0} (1-{\nu_x-\nu_{x0}\over\nu_{x0}}+{1\over\nu_{xo}}
Re \sum_{m\ne 0} m f_m \exp(-im\theta_x)) \cr
{\beta_x-\beta_{x0}\over\beta_{x0}} &= -{\nu_x-\nu_{x0}\over\nu_{x0}} +
{1\over\nu_{x0}} Re \sum_{m\ne 0} m f_m \exp(-im\theta_x) \cr}
\eqno{\hbox{(5-4)}}$$
The results for $(\beta_x-\beta_{x0})/\beta_{x0}$ is then
$${\beta_x-\beta_{x0}\over\beta_{x0}} = - {\nu_x-\nu_{x0}\over\nu_{x0}} -
\sum_{m\ne 0} {2\over m-2\nu_{x0}} d_x Re (b_m \exp(-im\theta_x-i\delta_{1x}))
\eqno{\hbox{(5-5)}}$$

If one assumes that the harmonic $m\simeq 2\nu_{x0}$ dominate then the
maximum change in $\beta_x$ may be approximated by
$$\eqalign{
\left|{\beta_x-\beta_{x0}\over\beta_{x0}}\right|_{\rm max} &=
\left|{\nu_x-\nu_{x0}\over\nu_{x0}}\right| + {2|d_x||b_m|\over |m-2\nu_{x0}|}\cr
m &\simeq 2\nu_{x0} \cr} \eqno{\hbox{(5-6)}}$$
$\nu_x$ is given by Eq. (4-10).

The results for $\beta_y$ may be found by replacing each parameter with the
corresponding parameter for the $\nu_y$ mode.

\section{Small Accelerator Results}

All the final results found in this paper will also hold for small accelerators
where the exact equations of motion have to be used.  The exact linear
equations have the form [1]
$${dx_i\over ds} = \sum A_{ij} x_j, \eqno{\hbox{(6-1)}}$$
where $i,j$ for from 1 to 4 and the $x_i$ are the coordinates $x$, $p_x$,
$y$, $p_y$.  For large accelerators $p_x\simeq dx/ds$, $p_y\simeq dy/ds$,
$A_{11}=A_{22}=A_{33}=A_{44}=0$, and $A_{12}=A_{34}=1$.  The $A_{ij}$ for
the exact equations are given in reference 1.  In particular
$$\eqalign{
A_{12} &= {(1+x/\rho)(1-p_y^2)\over p_s^3}, \quad A_{34} =
{(1+x/\rho)(1-p_x^2)\over p_s^3} \cr
A_{13} &= 0 , \qquad A_{14} = {(1+x/\rho)p_xp_y\over p_s^3} \cr
p_s &= \left\{ 1-p_x^2-p_y^2 \right\}^{1\over 2} \cr} \eqno{\hbox{(6-2)}}$$
where the right hand side in the equation for $A_{ij}$ are evaluated on the
closed orbit.

The linear differential equations for $\eta_x$ and $\eta_y$, $\eta_x=
x/\beta_{x0}^{1\over}$, $\eta_y=\nu/\beta_{y0}^{1\over 2}$ can be found [1]
as
$$\eqalign{
{d^2\over d\theta_x^2} \eta_x+\nu_{x0}^2 \eta_x &= f_x \cr
{d^2\over d\theta_y^2} \eta_y+\nu_{y0}^2 \eta_y &= f_y \cr
f_x &= {\nu_{x0}^2\beta_{x0}^{3/2}\over A_{12}} (1+x/\rho)\Delta B_y \cr
f_y &= -{\nu_{yo}^2\beta_{y0}^{3/2}\over A_{34}} (1+x/\rho)\Delta B_x \cr
d\theta_x &= A_{12} ds/\nu_{x0}\beta_{x0}, \ \ d\theta_y=A_{34} ds/\nu_{y0}
\beta_{y0} \cr} \eqno{\hbox{(6-3)}}$$
$\Delta B_x$, $\Delta B_y$ are the perturbing fields given by Eqs. (2-1).  
For small accelerators, in order for Eqs. (6-3) to be valid, one also
requires that the perturbing fields do not shift the closed orbit, or
$\Delta B_x=\Delta B_y=0$ on the closed orbit.  If the closed orbit is
shifted by the perturbing field, then the non--linear kinematic terms,
the terms which do not explicitly depend on the field, will generate 
additional linear terms.  $\nu_{x0}$, $\nu_{y0}$, $\beta_{x0}$, $\beta_{y0}$
are the tune and beta functions of the unperturbed accelerator.

Comparing Eqs. (6-3) with the corresponding equations for large accelerators,
Eqs. (2-2) one notes that $f_x$ and $f_y$ for the small accelerator have 
the additional factors of $1/A_{12}$ and $1/A_{34}$ respectively.
Although the perturbation terms in Eqs. (6-3) now have the extra factors
$1/A_{12}$ and $1/A_{34}$, these factors disappear in the final results
when in the relevant integrals one goes from the variables $\theta_x$
or $\theta_y$ to the variable $s$ according to Eqs. (6-3).  Using Eqs. (6-3)
one can go through the derivations and show that the final results for the
tune, growth rates and beta functions are valid for both large and
small accelerators.

One thing that remains to be done is to derive Eq. (5-1),
$${1\over\beta_x}={1\over\nu_{x0}\beta_{x0}} Im {d\over d\theta_x} ln\ x
\eqno{\hbox{(6-4)}}$$
which allows $\beta_x$ to be computed from the solutions for $\eta_x$,
$\eta_y$.

The beta functions $\beta_x$, $\beta_y$ for linearly coupled motion may
be defined by going to the coordinate system where the near coordinates
$\mu$, $p_u$, $v$, $p_v$ are uncoupled.  If the $\mu$, $p_u$ motion
goes into the $x$, $p_x$ motion when the coupling goes to zero, then
the beta function of the uncoupled $\mu$, $p_u$ motion will be called
$\beta_x$.  The solution of the equations of motion for $\mu$, $p_u$
may be related to $\beta_x$ by [1]
$$\eqalign{
\mu &= C \beta_2^{1\over 2} \exp (i\psi_x) \cr
p_u &= C \beta_x^{-{1\over 2}} \exp(-\alpha_x+i)\exp(i\psi_x) \cr} 
\eqno{\hbox{(6-5)}}$$
$C$ is a normalization constant.  $x$, $p_x$ and $\mu$, $p_u$ are related
by the decoupling matrix, $R$ [3]
$$\eqalign{
\pmatrix{ x \cr p_x \cr y \cr p_y \cr} &= R \pmatrix{ u \cr p_u \cr v \cr  
p_v \cr} \cr
R &= \pmatrix{ \cos\phi I & D\sin\phi \cr -D^{-1} \sin\phi & \cos\phi I 
\cr} \cr} \eqno{\hbox{(6-6)}}$$
$I$ is the $2\times 2$ identity matrix and $D$ is a $2\times 2$ matrix and
$|D|=1$.  One then finds
$$\eqalign{
x &= C \beta_x^{1\over 2} \exp(i\psi_x) \cr
p_x &= C \beta_x^{-{1\over 2}} \exp(i\psi_x) \cr} \eqno{\hbox{(6-7)}}$$
From Eq. (6-7) one can relate $\beta_x$ to $x$, $p_x$ solutions by
$${1\over\beta_x} = Im (p_x/x) \eqno{\hbox{(6-8)}}$$
where $Im$ is the imaginary part.  $p_x$ may be eliminated by using
Eq. (6-1)
$$p_x = {1\over A_{12}} ({dx\over ds} - A_{11} x-A_{13} y-A_{14} p_y)$$
which gives
$${1\over\beta} = Im [{1\over x} {1\over A_{12}} ({dx\over ds}-A_{11} x -
A_{13} y -A_{14} p_y)] \eqno{\hbox{(6-9)}}$$

In the large accelerator approximation, $A_{12}=1$ and $A_{13}=A_{14}=0$
and (6-9) gives
$${1\over\beta} = {Im\over\nu_{x0}\beta_{x0}}{d\over d\theta_x} \log x
\eqno{\hbox{(6-10)}}$$
For small accelerators, one has (see reference 1)
$$\eqalign{
A_{13} &= {\partial\over\partial y} [{(1+x/\rho)p_x\over p_s}]=0 \cr
A_{14} &= {\partial\over\partial p_y} [{(1+x/\rho)p_x\over p_s}] =
(1+x/\rho) {p_xp_y\over p_s^3} \cr
p_s &= [-1-p_x^2-p_y^2]^{1/2} \cr} \eqno{\hbox{(6-11)}}$$
Thus if $\Delta B_x=0$ on the closed orbit, so that $p_y=0$ on the closed
orbit then $A_{14}=0$ for small accelerators too.  Since $A_{12}ds=
\nu_{x0}\beta_{x0}d\theta_x$, one gets Eq. (6-10) again for small
accelerators provided $\Delta B_x=0$ on the unperturbed closed orbit.

\section{Comments on the Results}

Others have worked on this subject and there is an overlap between the
contents of this paper and their work.  These previous papers (4 to 12)
give results for the stopband width and for the growth rate inside the
stopband.

The results in this paper include the following:

\enumerate

\itm Results for the tune in the stable region near an edge of the stopband.
The results show that as $\nu_{x0}$, $\nu_{y0}$ approach the edge of the
stopband, the tunes of the two normal modes $\nu_x$ and $\nu_y$ begin to
change rapidly and when $\nu_{x0}$, $\nu_{y0}$ reach the stopband edge
then $\nu_x$ and $\nu_y$ lie on the resonance line $\nu_x+\nu_y=q$.
These final values of $\nu_x$, $\nu_y$, when $\nu_{x0}$, $\nu_{y0}$ reach
the stopband edge, are approached like $\epsilon^{1\over 2}$, where 
$\epsilon$ is the distance from $\nu_{x0}$, $\nu_{y0}$ to the stopband
edge.

\itm Results for the beta functions of the normal modes, $\beta_x$, $\beta_y$,
in the stable region near the edge of a stopband.  The results show that
$\beta_x$, $\beta_y$ do not become infinite when $\nu_{x0}$, $\nu_{y0}$
approach the stopband edge, unless $\nu_{x0}$, $\nu_{y0}$ are near the
half integer resonances $v_x=m/2$, or $v_y=n/2$, $m$ and $n$ being
integers.

\itm Results for the 2 solutions of the equations of motion in the stable
region near a stopband edge and in the unstable region.

\itm The above results hold also for small accelerators, where the exact
equations of motion have to be used and the large accelerator approximation
is not valid.  For small accelerators, one needs the restriction that the
perturbing field gradients do not shift the closed orbit.

\endenumerate

\nosechead{References}

\enumerate

\itm G. Parzen, Linear orbit parameters for the exact equations of motion,
BNL Report, BNL--60090 (1994).

\itm G. Parzen, Particle motion in the stable region near the edge of a
linear half--integer resonance, BNL report, BNL--62036 (1995).

\itm D. Edwards and L. Teng, Parameterization of linear coupled motion in
periodic systems, Proc. 1973 IEEE PAC, p. 885 (1973).

\itm P.A. Sturrock, Static and dynamic electron optics, Cambridge Univ.
Press, London (1955).

\itm R. Hagedom, Stability and amplitude ranges of two--dimensional
non--linear oscillations with a periodic hamiltonian, CERN 57-1 (1957).

\itm A. Schoch, Theory of linear and non-linear perturbations of betatron
oscillations in an alternating gradient synchrotron, CERN 57-21 (1957).

\itm E.D. Courant and H.S. Snyder, Theory of the alternating gradient
synchrotron, Ann. Phys. \ul{3}, 1 (1958).

\itm A.A. Kolomensky and A.N. Lebedev, Theory of cyclic accelerators,
North Holland Publishing Co. (1966).

\itm G. Ripken, Untersuchugen zur Strahlf\"uhrung und Stabsilitat der
Teilchen bewegung in Beschleunigern und Storage--Ringes unter stronger
Ber\'uchsichtiegung einer Kopplung der Betatron schwingungen, DESY R1-7014
(1970).

\itm W.P. Lysenko, Nonlinear betatron oscillations, Particle Accelerators
\ul{5}, 1 (1973).

\itm G. Guignard, The general theory of all sum and difference resonances
in three--dimensional magnetic field in a synchrotron, CERN 76-06 (1976).

\itm G. Guignard, A general treatment of resonances in accelerators,
CERN 78-11 (1978).

\itm H. Wiedenmann, Particle Accelerator Physics II, Springer--Verlag,
New York (1995).

\endenumerate

\bye